\documentclass[twocolumn,nopacs,floatfix,amsmath,nofootinbib,amssymb,preprintnumbers,floatfix]{revtex4}
\usepackage{longtable,lscape,booktabs}
\usepackage{txfonts}
\usepackage{overpic}
\usepackage{amssymb}
\usepackage{indentfirst}
\usepackage{feynmf}
\usepackage{slashed}
\usepackage{cases}
\usepackage{color}
\usepackage{multirow}
\usepackage{ulem}
\usepackage{graphicx,color,dcolumn,bm}
\usepackage[colorlinks,
            citecolor=blue,
            anchorcolor=red,
            menucolor=red,
            linkcolor=red,
            filecolor=red,
            runcolor=red,
            urlcolor=blue,
            frenchlinks=red]{hyperref}

\begin{document}

\title{The $Z_{cs}$ states and the mixture of hadronic molecule and diquark-anti-diquark components within effective field theory}

\author{Ze-Hua Cao$^{3}$}
\author{Wei He$^{1,2}$}
\author{Zhi-Feng Sun$^{1,2,3,4}$}\email[Corresponding Author: ]{sunzf@lzu.edu.cn}

\affiliation
{
$^1$School of Physical Science and Technology, Lanzhou University, Lanzhou 730000, China\\
$^2$Research Center for Hadron and CSR Physics, Lanzhou University and Institute of Modern Physics of CAS, Lanzhou 730000, China\\
$^3$Lanzhou Center for Theoretical Physics, Key Laboratory of Theoretical Physics of Gansu Province, Lanzhou University, Lanzhou, Gansu 730000, China\\
$^4$Frontiers Science Center for Rare Isotopes, Lanzhou University, Lanzhou, Gansu 730000, China
}

\date{\today}

\begin{abstract}
In this work, we construct the Lagrangian describing meson-diquark interaction, such that the diquark-anti-diquark component as well as the molecular component is introduced when studying the $Z_{cs}$ states. In this way, the problem is solved that if only considering the $\bar{D}^{(*)}D_s^{(*)}$ components, the potentials are suppressed by OZI rule. Through solving the Bethe-Salpeter equation, we find that the $Z_{cs}(4000)^+$ can be explained as the mixture of $\bar{D}^{*0}D_s^+$ and $\bar{A}_{cs}S_{cu}$ components. Besides, for the $\bar{D}^{*0}D_s^{*+}/\bar{A}_{cs}A_{cu}$ system, the pole of $4208\pm 13i$ MeV on the second Riemann sheet is predicted, whose mass agrees with that of $Z_{cs}(4220)^+$ while the width is much smaller than $Z_{cs}(4220)^+$. Due to the large error of the $Z_{cs}(4220)^+$'s width, further measurements are expected. In addition, several other poles of different spins are predicted.
\end{abstract}

\maketitle

\section{Introduction}
A series of charmonium-like states have been discovered, since the observation of $X(3872)$ by Belle collaboration in 2003 \cite{Choi:2003ue}. 
Besides, searching for the $Z_{cs}$ states composed of $c$, $\bar{c}$, $\bar{s}$ and $q$ is an important topic as well. In 2020, BESIII reports new results in the $e^+e^-\to K^+ (D_s^-D^{*0}+D_s^{*-}D^0)$ process.  An excess over the known contributions of the conventional charmed mesons is observed near the $D_s^-D^{*0}$ and $D_s^{*-}D^0$ mass thresholds in the $K^+$ recoil-mass spectrum for events collected at $\sqrt{s}= 4.681$ GeV \cite{BESIII:2020qkh}. The corresponding mass and width are determined as
\begin{eqnarray}
m=3982.5^{+1.8}_{-2.6}\pm 2.1\ {\rm MeV}, \ \Gamma=12.8^{+5.3}_{-4.4}\pm 3.0\ {\rm MeV}.
\end{eqnarray}
Soon after that, LHCb collaboration observed two exotic states decaying into $J/\psi K^+$ final state, which are named as $Z_{cs}(4000)^+$ and $Z_{cs}(4220)^+$ \cite{LHCb:2021uow}. Their masses and the widths are
\begin{eqnarray}
m_{Z_{cs}(4000)^+}&=&4003\pm 6^{+4}_{-14}\ {\rm MeV}, \\
\Gamma_{Z_{cs}(4000)^+}&=&131\pm 15\pm 26\ {\rm MeV},\\
m_{Z_{cs}(4220)^+}&=&4216\pm 24^{+43}_{-30}\ {\rm MeV}, \\
\Gamma_{Z_{cs}(4220)^+}&=&233\pm 52^{+97}_{-73}\ {\rm MeV}.
\end{eqnarray}
Very recently, the evidence of $Z_{cs}(3985)^0$ was found by BESIII near the thresholds of $D_s^-D^{*+}$ and $D_s^{*-}D^+$ production in the $K_S^0$ recoil-mass spectrum \cite{BESIII:2022qzr}. Since its mass and width are close to those of $Z_{cs}(3985)^+$, they should be the isospin partners. However, $Z_{cs}(3985)$ and $Z_{cs}(4000)$ seem not to be the same states, due to the different masses and widths. 

These exciting discoveries enrich our understanding of the spectrum of hadronic states, and much attention is paid in this field. Before the observations, the authors in Ref. \cite{Dias:2013qga} predicted the decay width of a charged state near the $D_s\bar{D}^*/D_s^*\bar{D}$ threshold. Ref. \cite{Voloshin:2019ilw} pointed out that it would be natural to expect the existence of $Z_{cs}$ states to decay into $\eta_c K$ and $J/\psi K$. After the observations, in Refs. \cite{Meng:2020ihj,Yang:2020nrt,Wan:2020oxt,Ortega:2021enc,Chen:2021erj,Meng:2021rdg,Yan:2021tcp,Xu:2020evn,Wang:2020htx,Shi:2021jyr,Maiani:2021tri,Yang:2021zhe,Giron:2021sla,Wang:2020iqt,Karliner:2021qok,Han:2022fup,Chen:2021uou,Guo:2020vmu,Baru:2021ddn,Albuquerque:2021tqd,Zhai:2022ied,Ozdem:2021hka,Wu:2021ezz}, the $Z_{cs}$ states are studied in the pictures of molecular,  compact tetraquark or the mixture of them. For other theoretical works, see Refs. \cite{Wang:2020dmv,Wang:2020kej,Ge:2021sdq,Ikeno:2020mra,Du:2022jjv,Zhu:2021vtd,Chen:2022yev}.

Since the masses of the heavy quarks are much larger than those of the light quarks, the heavy quarks can be seen as spectators in hadrons. From this point of view, the light-meson exchange is dominant in the charmed and anti-charmed mesons interactions. In this case, if only considering the $\bar{D}^{(*)}D_s^{(*)}$ hadronic molecule component when studying the observed $Z_{cs}$ states, the scattering processes corresponding to the effective potentials are suppressed by the Okubo-Zweig-Iizuka (OZI) rule \cite{Okubo:1963fa,Zweig,Iizuka:1966fk}. Such suppressed potentials may not be enough to form a particle. In order to solve this issue, in present work, we introduce the effective field theory for both mesons and diquarks, in which way, the diquark-anti-diquark component is introduced for the $Z_{cs}$ states. Next, we give a short review of the effective field theory.

As is well known, chiral perturbation theory (ChPT) is a powerful tool to study the phenomenon in the low energy region of quantum chromodynamics (QCD).  It is based on the spontaneously broken approximate chiral symmetry of QCD. The pions, kaons and the eta can be identified with the Goldstone bosons of chiral symmetry breaking. However, in the higher energy region, ChPT may not be applicable. One simple way is to incorporate vector mesons in this energy region. In order to achieve this, there are various schemes \cite{Meissner:1987ge,Birse:1996hd,Harada:2003jx}, such as the hidden local symmetry (HLS) approach, the matter field method, the antisymmetric tensor field method and the massive Yang-Mills field method.

In this work, we focus on the first one, i.e., the HLS approach. In this approach \cite{Furui:1995bj}, an artificial local symmetry is introduced into the nonlinear sigma model by the choice of field variables. The vector mesons are then introduced as gauge bosons for this symmetry. As stressed in Refs. \cite{Georgi:1989gp,Georgi:1989xy}, the additional local symmetry has no physics associated with it, and it can be removed by fixing the gauge. In the unitary gauge, the symmetry reduces to a nonlinear realisation of chiral symmetry, under which the vector fields transform inhomogeneously. In this work, we treat the mesons and diquarks as point-like particles, and extend the HLS to construct the Lagrangian describing the interactions of them. For the papers containing the Lagrangians of light mesons and diquarks, one can refer to the Refs. \cite{Ebert:1995fp,Harada:2019udr,Kim:2020imk,Kim:2021ywp}. After obtaining the potentials of the $Z_{cs}$ states within the mixture of hadronic molecule and diquark-antidiquark components, we solve the Bethe-Salpeter equation of the on-shell factorized form. For this step, we follow the chiral unitary approach, which is successfully used in studying the molecular states, see, for instance, the Refs. \cite{Oller:2000ma,Geng:2008gx,Molina:2010tx,Wu:2010jy}. 

The paper is organized as follows. In Sec. II, we discuss the necessity of diquark-anti-diquark component. In Sec. III, we introduce the construction of the Lagrangian incorporating mesons and diquarks under the HLS method. In Sec. IV and V, the effective potentials are calculated and the Bethe-Salpeter equation of the on-shell factorized form is given. After that, we show the result in Sec. VI. Finally, a short summary is given.

\section{The necessity of diquark-antidiquark component}

\begin{figure}
\centering
\vspace{-1.2cm}
\setlength{\abovecaptionskip}{-5.3cm} 
\includegraphics[width=1.05\linewidth]{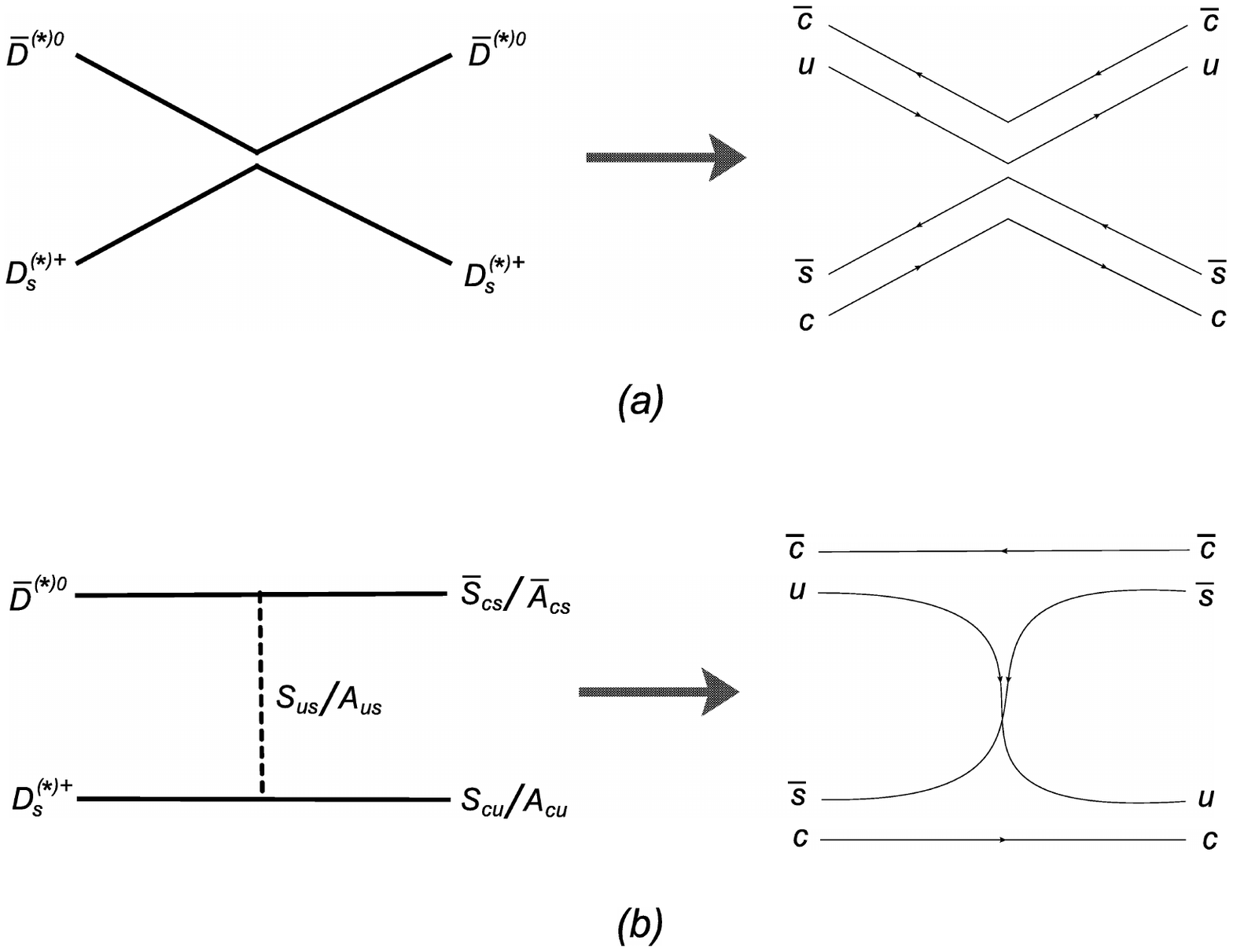}
\caption{The diagrams of mesons into mesons and mesons into diquark-anti-diquark.\label{fig1}}
\end{figure}

If we only consider the meson-meson interactions for $Z_{cs}$ states, the corresponding scattering processes are OZI suppressed as shown in Fig. \ref{fig1}(a). Note that the heavy quarks here are seen as spectators.

However, if we introduce the diquark-anti-diquark component, the issue above is naturally solved. The reason is that the meson-meson component into the diquark-anti-diquark component could occur via the exchange of a light diquark (see Fig. \ref{fig1}(b)). In this way, the corresponding process is OZI allowed.

There are several types of diquarks. Each of the quarks is a color triplet, such that a diquark can be either a color anti-triplet or a color sextet. However, only the color anti-triplet of the diquarks contributes since there is a repulsion between the quarks in the color sextet and an attraction in the color anti-triplet.

At the meson-diquark level, the diquark fields are depicted as
\begin{eqnarray}
S^a&=&\left(\begin{array}{ccc}
     0&S_{ud}&S_{us}\\
     -S_{ud}&0&S_{ds}\\
     -S_{us}&-S_{ds}&0
    \end{array}\right)^a,\\
A_\mu^a&=&\left(\begin{array}{ccc}
     A_{uu}&\frac{1}{\sqrt{2}}A_{ud}&\frac{1}{\sqrt{2}}A_{us}\\
     \frac{1}{\sqrt{2}}A_{ud}&A_{dd}&\frac{1}{\sqrt{2}}A_{ds}\\
     \frac{1}{\sqrt{2}}A_{us}&\frac{1}{\sqrt{2}}A_{ds}&A_{ss}
    \end{array}\right)_\mu^a,\\
S_c^a&=&\left(\begin{array}{ccc}
     S_{cu}&S_{cd}&S_{cs}
    \end{array}\right)^a,\\
A_{c\mu}^a&=&\left(\begin{array}{ccc}
     A_{cu}&A_{cd}&A_{cs}
    \end{array}\right)_\mu^a,
\end{eqnarray}
where $S^a$ is the light scalar diquark, $A_\mu^a$ the light axial vector diquark, $S_c^a$ the charmed scalar diquark, and $A_{c\mu}^a$ the charmed axial vector diquark. The superscript $a=1,2,3$ is the color index.

\section{The Lagrangian}
In this work, we extend the HLS to the meson-diquark interaction sector, considering as well the chiral symmetry, parity, and charge conjugation. The constructed Lagrangian is shown below
\begin{eqnarray}
\mathcal{L}&=&e_1(iPD_\mu SA_c^{\mu\dag}-iA_c^\mu D_\mu S^\dag P^\dag)\nonumber\\&&+e_2(iPA_\mu D^\mu S_c^\dag-iD^\mu S_cA^\dag_\mu P^\dag)\nonumber\\
&&+e_3(\epsilon^{\mu\nu\alpha\beta}PA_{\mu\nu}A_{c\alpha\beta}^\dag+\epsilon^{\mu\nu\alpha_\beta}A_{c\alpha\beta}A^\dag_{\mu\nu}P^\dag)\nonumber\\
&&+e_4(iP^*_\mu D^\mu SS_c^\dag-iS_cD^\mu S^\dag P^{*\dag}_\mu)\nonumber\\
&&+e_5(\epsilon^{\mu\nu\alpha\beta}P_\mu^*D_\nu SA_{c\alpha\beta}^\dag+\epsilon^{\mu\nu\alpha\beta}A_{c\alpha\beta}D_\nu S^\dag P_\mu^{*\dag})\nonumber\\
&&+e_6(\epsilon^{\mu\nu\alpha\beta}P_\mu^*A_{\nu\alpha}D_\beta S_c^\dag+\epsilon^{\mu\nu\alpha\beta}D_\beta S_cA_{\nu\alpha}^\dag P_\mu^{*\dag})\nonumber\\
&&+e_7(iP^*_\mu A^{\mu\nu}A_{c\nu}^\dag-iA_{c\nu}A^{\mu\nu\dag}P^*_\mu)\nonumber\\
&&+e_8(iP^*_\mu A_\nu A_c^{\mu\nu\dag}-iA_c^{\mu\nu}A_\nu^\dag P^{*\dag}_\mu)\nonumber\\
&&+e_9(iP^*_{\mu\nu}A^\mu A_c^{\nu\dag}-iA_c^\nu A^{\mu\dag}P^{*\dag}_{\mu\nu}),\label{eqLagrangian}
\end{eqnarray}
where 
\begin{eqnarray}
P&=&(D^0,D^+,D_s^+),\
P^*_\tau =(D^{*0},D^{*+},D_s^{*+})_\tau,\\
D_\mu P&=&\partial_\mu P+iP\alpha^\dag _{\| \mu}=\partial_\mu P+iP\alpha_{\| \mu},\\
D_\mu P^*_\tau&=&\partial_\mu P^*_\tau+iP^*_\tau\alpha^\dag _{\| \mu}=\partial_\mu P^*_\tau+iP^*_\tau\alpha_{\| \mu},\\
\alpha_{\bot\mu}&=&(\partial_\mu \xi_R\xi_R^\dag-\partial_\mu \xi_L\xi_L^\dag)/(2i),\\
\alpha_{\|\mu}&=&(\partial_\mu \xi_R\xi_R^\dag+\partial_\mu \xi_L\xi_L^\dag)/(2i),\\
\hat{\alpha}_{\bot\mu}&=&(D_\mu \xi_R\xi_R^\dag-D_\mu \xi_L\xi_L^\dag)/(2i),\\
\hat{\alpha}_{\|\mu}&=&(D_\mu \xi_R\xi_R^\dag+D_\mu \xi_L\xi_L^\dag)/(2i),\\
\xi_L&=&e^{i\sigma/F_\sigma}e^{-im/(2F_\pi)},\
\xi_R=e^{i\sigma/F_\sigma}e^{im/(2F_\pi)},\\
A^a_{\mu\nu}&=&D_\mu A^a_\nu-D_\nu A^a_\mu,\
A^a_{c\mu\nu}=D_\mu A^a_{c\nu}-D_\nu A^a_{c\mu},\\
D_\mu A_\nu^a&=&\partial_\mu A_\nu^a-iV_\mu A_\nu^a-iA_\nu^a V_\mu^T,\\
D_\mu S^a&=&\partial_\mu S^a-iV_\mu S^a-iS^aV_\mu^T,\\
D_\mu A_{c\nu}^a&=&\partial_\mu A_{c\nu}^a-iA_{c\nu}^a\alpha_{\|\mu}^T,\
D_\mu S_c^a=\partial_\mu S_c^a-iS_c^a\alpha_{\|\mu}^T
\end{eqnarray}
with the pseudoscalar meson fields
\begin{eqnarray}
\Phi&=&\left(\begin{array}{ccc}
\frac{\sqrt{3}\pi^0+\eta_8+\sqrt{2}\eta_0}{\sqrt{3}}&\sqrt{2}\pi^+ &\sqrt{2}K^{+}\\
\sqrt{2}\pi^-&\frac{-\sqrt{3}\pi^0+\eta_8+\sqrt{2}\eta_0}{\sqrt{3}}&\sqrt{2}K^{0}\\
\sqrt{2}K^{-}&\sqrt{2}\bar{K}^{0}&\frac{-2\eta_8+\sqrt{2}\eta_0}{\sqrt{3}}
\end{array}\right),
\end{eqnarray}
and the gauge boson fields
\begin{eqnarray}
V_\mu&=&\frac{g_V}{\sqrt{2}}\left(\begin{array}{ccc}
\frac{1}{\sqrt{2}}(\rho^0+\omega)&\rho^+ &K^{*+}\\
\rho^-&-\frac{1}{\sqrt{2}}(\rho^0-\omega)&K^{*0}\\
K^{*-}&\bar{K}^{*0}&\phi
\end{array}\right)_\mu.
\end{eqnarray}
The pion decay constant $F_\pi=93$ MeV. In the unitary gauge, i.e., $\sigma=0$, we have $\xi_L=\xi_R=e^{-i\Phi/(2F_\pi)}$. The $e_i$ ($i=1,2,...,9$) are the coupling constants.

\section{The effective potentials}
With the constructed Lagrangian in Eq. \eqref{eqLagrangian}, we calculate the potentials of two mesons into diquark-anti-diquark, which are shown in the following
\begin{eqnarray}
V^{\bar{D}^{*0}D_s^+\to \bar{A}_{cs}^aS_{cu}^a}&=&-\frac{e_2\sqrt{m_{S_{cu}}m_{D_s^+}m_{\bar{A}_{cs}}m_{\bar{D}^{*0}}}}{2m_{A_{us}}^2}\left(e_8\frac{s-m_{\bar{A}_{cs}}^2-m_{S_{cu}}^2}{2}\right.\nonumber\\
&&\left. +e_9\frac{m_{\bar{D}^{*0}}^2+m_{S_{cu}}^2-u}{2}\right)\vec{\epsilon}_1\cdot \vec{\epsilon}_3^\dag,\\
V^{\bar{D}^0D_s^{*+}\to \bar{S}_{cs}^aA_{cu}^a}&=&-\frac{e_2\sqrt{m_{\bar{D}^0}m_{\bar{S}_{cs}}m_{A_{cu}}m_{D_s^{*+}}}}{2m_{A_{us}}^2} \left(e_8\frac{s-m_{\bar{S}_{cs}}^2-m_{A_{cu}}^2}{2}\right.\nonumber\\
&&\left.+e_9\frac{m_{D_s^{*+}}^2+m_{\bar{S}_{cs}}^2-u}{2}\right)\vec{\epsilon}_2\cdot \vec{\epsilon}^\dag_4,\\
V^{\bar{D}^{*0}D_s^{*+}\to \bar{A}_{cs}^aA_{cu}^a}&=&-\frac{1}{2}\sqrt{m_{\bar{D}^{*0}}m_{\bar{A}_{cs}}m_{A_{cu}}m_{D_s^{*+}}}\left(e_8^2\frac{s-m_{\bar{A}_{cs}}^2-m_{A_{cu}}^2}{2}\right.\nonumber\\
&&+e_8e_9\frac{m_{D_s^{*+}}^2+m_{\bar{A}_{cs}}^2+m_{\bar{D}^{*0}}^2+m_{A_{cu}}^2-2u}{2}\nonumber\\
&&\left.+e_9^2\frac{s-m_{\bar{D}^{*0}}^2-m_{D_s^{*+}}^2}{2}\right)\frac{1}{m_{A_{us}}^2}\vec{\epsilon}_1\cdot\vec{\epsilon}_3^\dag\vec{\epsilon}_2\cdot\vec{\epsilon}_4^\dag
\end{eqnarray}
with $s=(p_1+p_2)^2$ and $u=(p_1-p_4)^2$.
Here, other amplitudes are negligible due to the OZI suppressed processes and the very small momentums of the initial and final particles. 

Then we project the polarization vector products into different spin states
\begin{eqnarray}
\mathcal{P}(1)&=&\vec{\epsilon}_1\cdot \vec{\epsilon}_3^\dag=\vec{\epsilon}_2\cdot \vec{\epsilon}_4^\dag,\\
\mathcal{P}^\prime(0)&=&\frac{1}{3}\vec{\epsilon}_1\cdot \vec{\epsilon}_2 \vec{\epsilon}_3^\dag\cdot \vec{\epsilon}_4^\dag,\\
\mathcal{P}^\prime(1)&=&\frac{1}{2}(\vec{\epsilon}_1\cdot \vec{\epsilon}_3^\dag \vec{\epsilon}_2\cdot \vec{\epsilon}_4^\dag-\vec{\epsilon}_1\cdot \vec{\epsilon}_4^\dag \vec{\epsilon}_2\cdot \vec{\epsilon}_3^\dag),\\
\mathcal{P}^\prime(2)&=&\frac{1}{2}(\vec{\epsilon}_1\cdot \vec{\epsilon}_3^\dag \vec{\epsilon}_2\cdot \vec{\epsilon}_4^\dag+\vec{\epsilon}_1\cdot \vec{\epsilon}_4^\dag \vec{\epsilon}_2\cdot \vec{\epsilon}_3^\dag)\nonumber\\
&&-\frac{1}{3}\vec{\epsilon}_1\cdot \vec{\epsilon}_2 \vec{\epsilon}_3^\dag\cdot \vec{\epsilon}_4^\dag.
\end{eqnarray}

\section{The Bethe-Salpeter Equation in The On-shell Factorized form}
With the effective potentials obtained above, we utilize the Bethe-Salpeter equation of the on-shell factorized form to calculate the T-matrix considering the coupled-channel effect, i.e.,
\begin{eqnarray}
T&=&(I-VG)^{-1}V.
\end{eqnarray}
Here, $V$ is the matrix of the effective potential, and $G$ is the matrix of the two-particle loop function whose non-zero element has the form of
\begin{eqnarray}
G_{ii}&=&i\int \frac{d^4q}{(2\pi)^4}\frac{1}{q^2-m_{i1}^2+i\epsilon}\frac{1}{(P-q)^2-m_{i2}^2+i\epsilon},
\end{eqnarray}
where $P_\mu$ is the four momentum of the two particles, $q_\mu$ is the four momentum of one of the particles, $m_{i1}$ and $m_{i2}$ are the masses of the particles in the loop, $i$ is the label of the channel. The above loop integral is logarithmically divergent, which can be regularized by the three-momentum-cutoff. The expression of the cutoff-regularized loop function has been calculated in Ref. \cite{Oller:1998hw}, i.e.,
\begin{eqnarray}
G_{ii}&=&\frac{1}{32\pi^2}\left\{\frac{\nu}{s}\left[\log\frac{s-\Delta+\nu\sqrt{1+\frac{m_{i1}^2}{q_{max}^2}}}{-s+\Delta+\nu\sqrt{1+\frac{m_{i1}^2}{q_{max}^2}}}\right.\right.\nonumber\\
&&\left.+\log\frac{s+\Delta+\nu\sqrt{1+\frac{m_{i2}^2}{q_{max}^2}}}{-s-\Delta+\nu\sqrt{1+\frac{m_{i2}^2}{q_{max}^2}}}\right]-\frac{\Delta}{s}\log\frac{m_{i1}^2}{m_{i2}^2}\nonumber\\
&&+\frac{2\Delta}{s}\log\frac{1+\sqrt{1+\frac{m_{i1}^2}{q_{max}^2}}}{1+\sqrt{1+\frac{m_{i2}^2}{q_{max}^2}}}+\log\frac{m_{i1}^2m_{i2}^2}{q_{max}^4}\nonumber\\
&&\left.-2\log\left[\left(1+\sqrt{1+\frac{m_{i1}^2}{q_{max}^2}}\right)\left(1+\sqrt{1+\frac{m_{i2}^2}{q_{max}^2}}\right)
\right]\right\}\label{LoopFunction}
\end{eqnarray}
with $q_{max}$ the cutoff, $\Delta=m_{i2}^2-m_{i1}^2$, and $\nu=\sqrt{[s-(m_{i1}+m_{i2})^2][s-(m_{i1}-m_{i2})^2]}$. 

Eq. \eqref{LoopFunction} justifies on the first Riemann sheet, on which the bound state can be found. In order to look for the pole of resonance or virtual state, we need to extrapolate the loop function to the second Riemann sheet by a continuation via
\begin{eqnarray}
G_{ii}^{II}&=&G_{ii}^I+i\frac{\nu}{8\pi s}.
\end{eqnarray}
Since the amplitudes close to a pole behave like
\begin{eqnarray}
T_{ij}&=&\frac{g_ig_j}{s-s_R},
\end{eqnarray}
we can  calculate the couplings to each channel from the residue of the amplitudes.

\section{Numerical Results}
In the Lagrangian shown above, there are 9 unknown constants. To determine them, we approximately use the quark-pair-creation model. Their values are given in Tab. \ref{tab1}. The masses of diquarks are taken from Ref. \cite{Ferretti:2019zyh}.
\begin{table}
\caption{The values of the low energy constants in the Lagrangian.}\label{tab1}
\begin{tabular}{c|c|c|c|c}\toprule[1pt]
$e_1$ (GeV$^{-1}$)&$e_2$ (GeV$^{-1}$)&$e_3$ (GeV$^{-2}$)&$e_4$ (GeV$^{-1}$)&$e_5$ (GeV$^{-2}$)\\\midrule[0.5pt]
-6.939&4.161&$\pm$1.520&4.039&$\pm$1.588\\\toprule[1pt]
$e_6$ (GeV$^{-2}$)&$e_7$ (GeV$^{-1}$)&$e_8$ (GeV$^{-1}$)&$e_9$ (GeV$^{-1}$)&\\\midrule[0.5pt]
$\pm$2.595&-2.840&-16.778&11.098&\\\bottomrule[1pt]
\end{tabular}
\end{table}
Because the relative phase between the amplitudes obtained from the Lagrangian and the quark-pair-creation model can not be completely fixed, the values of ($e_7$, $e_8$, $e_9$) are possibly chosen as ($2.840$ GeV$^{-1}$, $-11.098$ GeV$^{-1}$, $16.778$ GeV$^{-1}$) and ($-16.778$ GeV$^{-1}$, $-2.840$ GeV$^{-1}$, $-2.840$ GeV$^{-1}$), respectively, where we omit the unphysical ones. Next we use the values in Tab. \ref{tab1}, and discuss the other cases finally.

By solving the Bethe-Salpeter equation, we find the pole positions. 
In Tab. \ref{tab2}, we show the dependence of the pole positions on the cutoff. The spins of $\bar{D}^{*0}D_s^+/\bar{A}_{cs}S_{cu}$ and $\bar{D}^{0}D_s^{*+}/\bar{S}_{cs}A_{cu}$ systems are 1. In the case of $\bar{D}^{*0}D_s^{*+}/\bar{A}_{cs}A_{cu}$, the spin of the states can be 0, 1 and 2, since the pole positions are the same for all these spins.

For the $\bar{D}^{*0}D_s^+/\bar{A}_{cs}S_{cu}$ system, if the cutoff is chosen as $q_{max}=1385$ MeV, we get a resonance with the pole position at ($4013\pm 42i$) MeV on the second Riemann sheet. The mass of this resonance is consistent with that of $Z_{cs}(4000)^+$, and the width about $84$ MeV is close to the lower limit of the $Z_{cs}(4000)^+$ width. That is to say, $Z_{cs}(4000)^+$ is identified as the mixture of $\bar{D}^{*0}D_s^+$ and $\bar{A}_{cs}S_{cu}$ components. The module of the coupling to $\bar{D}^{*0}D_s^+$ ($\bar{A}_{cs}S_{cu}$) channel is calculated as 9.80 GeV (18.60 GeV), which means that the $\bar{A}_{cs}S_{cu}$ component is dominant.

For the $\bar{D}^{*0}D_s^{*+}/\bar{A}_{cs}A_{cu}$ system of spin 1, if choosing the cutoff as $1385$ MeV, a resonance at ($4208\pm 13i$) MeV on the second Riemann sheet is obtained with the modules of the couplings $|g_{\bar{D}^{*0}D_s^{*+}}|=4.88$ GeV and $|g_{\bar{A}_{cs}A_{cu}}|=7.62$ GeV. The mass is in agreement with the one of $Z_{cs}(4220)^+$, however, the width is much smaller than that of $Z_{cs}(4220)^+$. We suggest the experiments make further measurements, since the error of the width by LHCb is very large. Meanwhile, we predict a virtual state on the second Riemann sheet of $4106$ MeV, and a bound state on the first Riemann sheet of $4091$ MeV. The modules of the couplings of the virtual state and the bound state to $\bar{D}^{*0}D_s^{*+}$ ($\bar{A}_{cs}A_{cu}$) channel are 4.39 GeV (8.02 GeV) and 7.79 GeV (11.27 GeV), respectively. For spin 0 and 2, the obtained pole positions are the same as those of spin 1. 

For the $\bar{D}^{0}D_s^{*+}/\bar{S}_{cs}A_{cu}$ system, with the cutoff $q_{max}=1385$ MeV, there is a pole at ($4112\pm 66i$) MeV. The modules of the couplings to $\bar{D}^{0}D_s^{*+}$ and $\bar{S}_{cs}A_{cu}$ are 9.12 GeV and 16.27 GeV, respectively.

\begin{table}
\caption{The pole positions depending on the cutoff $q_{max}$ in the two meson loop function. RS-I and RS-II stand for the first and the second Riemann sheets, respectively. In the case of $\bar{D}^{*0}D_s^{*+}/\bar{A}_{cs}A_{cu}$, the spin of the states can be 0, 1 and 2. All the values are in the unit of MeV.}\label{tab2}
\begin{tabular}{cc|c|c|ccccccccc}\toprule[1pt]
$q_{max}$&&1360&1385&1410\\\midrule[0.5pt]
$\bar{D}^{*0}D_s^+/\bar{A}_{cs}S_{cu}$&RS-II&$4018\pm i46$&$4013\pm i42$&$4009\pm i39$\\\midrule[0.5pt]
&RS-II&$4209\pm i13$&$4208\pm i13$&$4207\pm i12$\\
$\bar{D}^{*0}D_s^{*+}/\bar{A}_{cs}A_{cu}$&RS-II&$4105$&$4106$&$4107$\\
&RS-I&$4088$&$4091$&$4093$\\
\midrule[0.5pt]
$\bar{D}^{0}D_s^{*+}/\bar{S}_{cs}A_{cu}$&RS-II&$4115\pm i69$&$4112\pm i66$&$4108\pm i64$\\\bottomrule[1pt]
\end{tabular}
\end{table}

As discussed above, there are two other choices of the coupling constants $e_7$, $e_8$ and $e_9$. However, if choosing these values, we can neither explain the $Z_{cs}(4000)^+$ and $Z_{cs}(4220)^+$ nor explain the $Z_{cs}(3985)$ observed by BESIII.

For the further experiments, as well as the $J/\psi K$ channel as measured by LHCb, the $J/\psi K^*$ channel may also be of importance, since the resonances of spin 0 and spin 2 would be observed in this channel.

\section{Summary}

In this work, applying the HLS approach to the meson-diquark sector, we construct the Lagrangian to describe the corresponding interactions. And we study the $Z_{cs}$ states as the mixture of the hadronic molecule and the diquark-anti-diquark components. It is in this way the problem is solved that if considering only the pure molecular component, the interaction is OZI suppressed. 

With the potentials obtained by using the Lagrangian, we solve the Bethe-Salpeter equation. We find three resonances of spin 1 with the pole positions on the second Riemann sheet of ($4013\pm 42i$) MeV, ($4208\pm 13i$) MeV, ($4112\pm 66i$) MeV for $\bar{D}^{*0}D_s^+/\bar{A}_{cs}S_{cu}$, $\bar{D}^{*0}D_s^{*+}/\bar{A}_{cs}A_{cu}$ and $\bar{D}^{0}D_s^{*+}/\bar{S}_{cs}A_{cu}$ systems, respectively. The first resonance is identified as the $Z_{cs}(4000)^+$, since the calculated mass agrees with the value by LHCb, and the obtained width is close to the lower limit of the LHCb data. The mass of the second resonance agrees with that of $Z_{cs}(4220)^+$, however, the width is much smaller than that of $Z_{cs}(4220)^+$. Due to the large error of the experimental data of $Z_{cs}(4220)^+$ width, further experiments are expected. There are as well a bound state and a virtual state with the masses around $4100$ MeV and spins $J=1$ for $\bar{D}^{*0}D_s^{*+}/\bar{A}_{cs}A_{cu}$ system. For spins 0 and 2 of $\bar{D}^{*0}D_s^{*+}/\bar{A}_{cs}A_{cu}$ system, the obtained pole positions are the same as those of the spin 1 case. For all these states, the dominant channels are the diquark-anti-diquark channels. Further experiments are expected to explore all these states. Besides the $J/\psi K$ channel, the $J/\psi K^*$ channel may also be important.

\section*{Acknowledgments}
This project is supported by the National Natural Science Foundation of China (NSFC) under Grants No. 11965016 and  11705069. We would like to thank Lanzhou Center for Theoretical Physics for the support to this project under the Grant No. 12047501 (NSFC). In this work, Ze-Hua Cao and Wei He contribute equally.

\bibliographystyle{plain}

\end{document}